\documentclass[11pt,eqs]{article}
\usepackage{latexsym}

\def\cleq{\setcounter{equation}{0}}

\makeatletter
\newcommand\xleftrightarrow[2][]{%
  \ext@arrow 9999{\longleftrightarrowfill@}{#1}{#2}}
\newcommand\longleftrightarrowfill@{%
  \arrowfill@\leftarrow\relbar\rightarrow}
\makeatother

\title{
T-duality as coordinates permutation in double space
\thanks{Work supported in part by
the Serbian Ministry of Education and Science, under contract No. 171031.}}
\author{B. Sazdovi\'c
\thanks{e-mail: sazdovic@ipb.ac.rs}\\
{\it Institute of Physics,}\\
{\it University of Belgrade,}\\
{\it 11001 Belgrade, P.O.Box 57, Serbia}}

\begin{document}
\maketitle
\begin{abstract}
We introduce the $2D$ dimensional double space with the
coordinates $Z^M= (x^\mu, y_\mu)$ which components are the
coordinates of initial space $x^\mu$ and its T-dual $y_\mu$.
We shall show that in this extended space the T-duality transformations can be realized simply by exchanging places
of some coordinates $x^a$,  along which we want to perform T-duality
and the corresponding dual coordinates $y_a$. In such approach it is
evident that T-duality leads to the physically equivalent theory and that complete set of T-duality transformations form
subgroup of the $2D$ permutation group.
So, in the double space we are able to represent the backgrounds of all T-dual theories in unified manner.
\end{abstract}

\vskip.3cm



\section{Introduction}

T-duality of the closed string has been investigated for a long time \cite{B,RV,GPR,AABL}. It transforms the theory of a string moving
in a toroidal background into the theory of a string moving in different toroidal background. Generally, one suppose that background has some continuous
isometries which leaves the action invariant. In suitable adopted coordinates, where the isometry acts as translation, it means that
background does not depend on some coordinates.

In the paper \cite{DS1} the new procedure for T-duality of the closed string, moving in $D$ dimensional weakly curved space,
has been considered. The generalized approach allows one to perform T-duality along coordinates on which the Kalb-Ramond field depends.
In that article T-duality transformations has been performed simultaneously along all coordinates.
It corresponds to   $ T^{full}=T^0\circ T^1\circ \dots \circ  T^{D-1}$ -duality relation with transformation of the coordinates
$y_\mu = y_\mu (x^\mu)$ connecting  the beginning and the end of the T-duality chain
\begin{eqnarray}\label{eq:chain2}
&&\Pi_{\pm \mu \nu},\,x^{\mu}
\mathrel{{\mathop{\mathop{\rightleftharpoons}^{\mathrm{T^1}}_{\mathrm{T_1}}}}}
\Pi_{1 \pm \mu \nu},\,x_1^{\mu}
\mathrel{{\mathop{\mathop{\rightleftharpoons}^{\mathrm{T^2}}_{\mathrm{T_2}}}}}
\Pi_{2 \pm \mu \nu},\,x_2^{\mu}
\mathrel{{\mathop{\mathop{\rightleftharpoons}^{\mathrm{T^3}}_{\mathrm{T_3}}}}}
\dots
\mathrel{{\mathop{\mathop{\rightleftharpoons}^{\mathrm{T^D}}_{\mathrm{T_D}}}}}
\Pi_{D \pm \mu \nu}={}^\star \Pi_{\pm \mu \nu},\,x_D^{\mu}=y_\mu  .
\end{eqnarray}
Here $\Pi_{i \pm \mu \nu}$ and $x^\mu_{i},\,(i=1,2,\cdots,D)$ are background fields and the coordinates of the corresponding configurations.
Applying the proposed procedure
$ T_{full}=T_0\circ T_1\circ \dots \circ  T_{D-1}$ to the T-dual theory one can obtain
the initial theory and the inverse duality relation
$x^\mu = x^\mu (y_\mu)$, connecting  the end  and the beginning of the T-duality chain.
For simplicity, in the article \cite{DS1} T-duality  has been performed along all directions.

The nontrivial extension of this approach, compared with the flat space case, is a source of closed string non-commutativity
\cite{Lust,ALLP,DNS}.
From the canonical point of view  considered in Ref.\cite{DNS}, there is similarity between open and closed string non-commutativity.
In both cases, the initial coordinates are given not only as a functions of some effective coordinates but
 as a linear combination of the effective coordinates and the effective momenta. It produces the nonzero Poisson brackets between coordinates.
In open  string case, such relation is solution of boundary conditions and
only endpoints, attached to $Dp$-brane, are non-commutative even in the flat space.

The closed string does not have endpoints and  the boundary conditions are satisfied automatically.
To understand the closed string non-commutativity, we should impose T-duality transformation laws and express the closed string coordinates of
T-dual theory in terms of the coordinates and momenta of original theory. Then  the  standard  Poisson brackets of the original theory induced nontrivial   Poisson brackets
between coordinates in the T-dual theory, which are proportional to the background fluxes times winding and  momenta numbers.
In order to obtain such feature, we need to introduce background fields which depend on the coordinates.
The simplest example is weakly curved background.

In $D$-dimensional space it is possible to perform T-duality along any subset of coordinates $x^a : T^a=T^0\circ T^1\circ \dots \circ  T^{d-1}$,
 and along corresponding T-dual  ones $y_a : T_a=T_0\circ T_1\circ \dots \circ  T_{d-1} \,\, ,(a=0,1,\cdots , d-1)$.
In the paper \cite{DNS2} this was done for the string moving in the weakly curved background.
For each case the T-dual actions, T-dual background fields and T-duality transformations has been obtained.
Let us stress that T-dualization ${\cal T}^a= T^a \circ T_a$ of the present paper in the $2D$ dimensional double space
contains two T-dualizations in terminology of Ref.\cite{DNS2}. In fact $D$ dimensional T-dualizations $ T^a$ and $T_a$ of the present paper are denoted
${\cal T}^a$ and ${\cal T}_a$ in Ref.\cite{DNS2}.

The introduction of the extended space of
double dimensions with coordinates $Z^M= (x^\mu, y_\mu)$ wil
help us to reproduce all results of Ref.\cite{DNS2} and offer
simple explanation for T-duality. In the present article we will demonstrate this for the flat background, while for the weakly curved background it
will be presented elsewhere \cite{SB}.
For example, T-duality
$T^{\mu_1}$ (along fixed coordinate $x^{\mu_1}$) and T-duality
$T_{\mu_1}$ (along corresponding  dual coordinate $y_{\mu_1}$) can be performed simply by
exchanging the places of the coordinates $x^{\mu_1}$ and
$y_{\mu_1}$ in the double space. It can be realized just
multiplying $Z^M$ with constant $2D \times 2D$ matrix. Similarly,
arbitrary T-duality $ {\cal T}^a=T^a \circ T_a $ can be realized
by exchanging the places of the coordinates $x^{\mu_1}, x^{\mu_2},
\cdots , x^{\mu_{d-1}}$ with the corresponding dual coordinates $y_{\mu_1}, y_{\mu_2}, \cdots ,
y_{\mu_{d-1}}$. From this explanation it is clear that T-duality
leads to the equivalent theory, because permutation of the
coordinates in double space can not change the physics.

Similar approach to T-duality, as a transformation in double space,  appeared long time ago
\cite{Duff}-\cite{WS2}. Interest in this topic emerged again with the articles \cite{Hull,Hull2}.
In the paper  \cite{Duff} the beginning and the nd of the chain (\ref{eq:chain2})
has been established. The relation of our approach and Ref.\cite{Hull} will be discussed in Sec.4.

The basic tools in our approach are T-duality  transformations connected
beginning and end of the chain. Rewriting these transformations in the double space we obtain the fundamental expression, where the generalized metric
relate derivatives of the extended coordinates. We will show that this expression is enough to find background fields from every nodes of the chain and T-duality
transformations between arbitrary nodes. In such a way we unify the beginning and all corresponding T-dual theories of the chain (\ref{eq:chain2}).

\cleq

\section{ T-duality in the double space}

Let us consider the closed bosonic string which propagates in D-dimensional space-time described by the action \cite{S}
\begin{equation}\label{eq:action0}
S[x] = \kappa \int_{\Sigma} d^2\xi\sqrt{-g}
\Big[\frac{1}{2}{g}^{\alpha\beta}G_{\mu\nu}[x]
+\frac{\epsilon^{\alpha\beta}}{\sqrt{-g}}B_{\mu\nu}[x]\Big]
\partial_{\alpha}x^{\mu}\partial_{\beta}x^{\nu},
\quad (\varepsilon^{01}=-1) \, .
\end{equation}

The string, with coordinates  $x^{\mu}(\xi),\ \mu=0,1,...,D-1$  is moving in the non-trivial background,
defined by the space-time metric $G_{\mu\nu}$ and the Kalb-Ramond field $B_{\mu\nu}$.
Here $g_{\alpha\beta}$ is intrinsic world-sheet metric and
the integration goes over two-dimensional world-sheet $\Sigma$
with coordinates $\xi^\alpha$ ($\xi^{0}=\tau,\ \xi^{1}=\sigma$).

The requirement of the world-sheet conformal invariance on the quantum level
leads to the space-time  equations of motion, which
in the lowest order in slope parameter $\alpha^\prime$,
for the constant dilaton field $\Phi=const$  are
\begin{equation}\label{eq:ste}
R_{\mu \nu} - \frac{1}{4} B_{\mu \rho \sigma}
B_{\nu}^{\ \rho \sigma}=0\, ,  \qquad
D_\rho B^{\rho}_{\ \mu \nu} = 0   \, .
\end{equation}
Here
$B_{\mu\nu\rho}=\partial_\mu B_{\nu\rho}
+\partial_\nu B_{\rho\mu}+\partial_\rho B_{\mu\nu}$
is the field strength of the field $B_{\mu \nu}$, and
$R_{\mu \nu}$ and $D_\mu$ are Ricci tensor and
covariant derivative with respect to space-time metric.

We will consider the simplest  solutions of (\ref{eq:ste})
\begin{equation}
G_{\mu\nu}=const,
\quad
B_{\mu\nu}  = const  \,  ,
\end{equation}
which satisfies the space-time equations of motion.

Choosing the conformal gauge $g_{\alpha\beta}=e^{2F}\eta_{\alpha\beta}$, and introducing  light-cone coordinates
$\xi^{\pm}=\frac{1}{2}(\tau\pm\sigma)$, $\partial_{\pm}= \partial_{\tau}\pm\partial_{\sigma}$,
the action (\ref{eq:action0}) can be rewritten in the form
\begin{equation}\label{eq:action1}
S  = \kappa \int_{\Sigma} d^2\xi\
\partial_{+}x^{\mu}
\Pi_{+\mu\nu}
\partial_{-}x^{\nu},
\end{equation}
where
\begin{eqnarray}
\Pi_{\pm\mu\nu} =
B_{\mu\nu} \pm\frac{1}{2}G_{\mu\nu}.
\end{eqnarray}


\subsection{Standard sigma-model T-duality}

Applying the T-dualization procedure on all the coordinates, we obtain the T-dual action \cite{DS1}
\begin{equation}\label{eq:dualna}
S[y]=
\kappa
\int d^{2}\xi\
\partial_{+}y_\mu
\,^\star \Pi_{+}^{\mu\nu}  \,
\partial_{-}y_\nu
=\,
\frac{\kappa^{2}}{2}
\int d^{2}\xi\
\partial_{+}y_\mu
\theta_{-}^{\mu\nu}
\partial_{-}y_\nu,
\end{equation}
where
\begin{eqnarray}
{\theta}^{\mu\nu}_{\pm}&\equiv&
-\frac{2}{\kappa}
(G^{-1}_{E}\Pi_{\pm}G^{-1})^{\mu\nu}=
{\theta}^{\mu\nu}\mp \frac{1}{\kappa}(G_{E}^{-1})^{\mu\nu} \,  .
\end{eqnarray}
Here we consider flat background and omit argument dependence of. Ref. \cite{DS1}.
The symmetric and antisymmetric parts of ${\theta}^{\mu\nu}_{\pm}$  are the inverse of the effective metric $G^E_{\mu\nu}$ and the non-commutativity parameter $\theta^{\mu\nu}$
\begin{eqnarray}
G^E_{\mu\nu} & \equiv  G_{\mu\nu}-4(BG^{-1}B)_{\mu\nu}, \qquad
\theta^{\mu\nu} & \equiv  -\frac{2}{\kappa} (G^{-1}_{E}BG^{-1})^{\mu\nu}.
\end{eqnarray}

Consequently, the T-dual background fields  are
\begin{equation}\label{eq:tdbf}
^\star G^{\mu\nu} =
(G_{E}^{-1})^{\mu\nu}, \quad
^\star B^{\mu\nu} =
\frac{\kappa}{2}
{\theta}^{\mu\nu}  \, .
\end{equation}
Note that the dual effective metric is just inverse of the
initial one
\begin{eqnarray}\label{eq:cdi1}
{}^\star G_E^{\mu\nu} \equiv {}^\star G^{\mu\nu}-4({}^\star B {}^\star G^{-1} {}^\star B)^{\mu\nu}= (G^{-1})^{\mu\nu}  \,  ,
\end{eqnarray}
and the following relations valid
\begin{equation}\label{eq:cdi2}
({}^\star B {}^\star G^{-1})^\mu{}_\nu = - (G^{-1} B)^\mu{}_\nu  \,  , \qquad
({}^\star G^{-1} {}^\star B)_\mu{}^\nu = -(B G^{-1})_\mu{}^\nu  \,  .
\end{equation}


\subsection{T-duality transformations}

The T-duality transformations between all initial coordinates $x^\mu$ and all dual coordinates $y_\mu$
of the closed string theory have been derived in ref.\cite{DS1}
\begin{eqnarray}\label{eq:xtdual}
\partial_{\pm}x^\mu \cong
-\kappa\theta^{\mu\nu}_{\pm}  \partial_{\pm} y_\nu
\,  , \qquad
\partial_{\pm}y_\mu\cong
-2\Pi_{\mp\mu\nu} \partial_{\pm} x^\nu .
\end{eqnarray}
They are inverse to one another. We omit argument dependence and $\beta^\pm_\mu$ functions because they  appear only in the weakly curved background.

We can put above T-duality transformations in a useful form,   where on the left hand side we put the terms with world-sheet antisymmetric tensor
$\varepsilon_\alpha{}^\beta$ (note that $\varepsilon_\pm{}^\pm = \pm 1$)
\begin{eqnarray}\label{eq:tdualc}
& \pm \partial_{\pm}y_\mu \cong  G^E_{\mu \nu}  \partial_{\pm}x^\nu
-2 [ B G^{-1}]_\mu{}^\nu \partial_{\pm}y_\nu \, ,  \nonumber \\
& \pm \partial_{\pm}x^\mu \cong 2 [G^{-1}  B ]^\mu {}_\nu \partial_{\pm}x^\nu
+(G^{-1})^{\mu \nu} \partial_{\pm}y_\nu \, .
\end{eqnarray}

Let us introduce the $2D$ dimensions double target space, which will play important role in the present article.
It contains both initial and T-dual coordinates
\begin{equation}\label{eq:escoor}
Z^M=\left (
\begin{array}{c}
 x^\mu  \\
y_\mu
\end{array}\right )\, .
\end{equation}
Here, as well as in Double field theory (for recent reviews see \cite{HZ}-\cite{BT}), all coordinates are doubled. It differs from approach of Ref.\cite{Hull} where only coordinates on the torus along which we perform T-dualization are doubled. The relation of our and that of Ref.\cite{Hull} will be discussed in Sec.4.

In terms of  double  space  coordinate we can rewrite the T-duality relations (\ref{eq:tdualc}) in the simple form
\begin{equation}\label{eq:tdual}
\partial_{\pm} Z^M \cong \pm \, \Omega^{MN} {\cal{H}}_{NK}  \,\partial_{\pm}Z^K \, ,
\end{equation}
where
\begin{equation}
\Omega^{MN}= \left (
\begin{array}{cc}
0 &  1 \\
1  & 0
\end{array}\right )\, ,
\end{equation}
is a constant symmetric matrix and we introduced so called generalized metric as
\begin{equation}\label{eq:gm}
{\cal{H}}_{MN}  = \left (
\begin{array}{cc}
 G^E_{\mu \nu}  &  -2 \,  B_{\mu\rho}  (G^{-1})^{\rho \nu}  \\
2 (G^{-1})^{\mu \rho} \,  B_{\rho \nu}  & (G^{-1})^{\mu \nu}
\end{array}\right )\, .
\end{equation}

It is easy to check that
\begin{equation}\label{eq:sodd}
{\cal{H}}^T  \Omega {\cal{H}} =\Omega \, .
\end{equation}
As noticed in Ref.\cite{Duff}, the relation  (\ref{eq:sodd}) shows that there exists manifest $O(D,D)$ symmetry.
In Double field theory   it is usual to call $\Omega^{MN}$ the $O(D,D)$ invariant metric and denote with $\eta^{MN}$.


\subsection{Equations of motions as consistency condition of T-duality relations}

It is well known that the equation of motion and the Bianchi identity
of the original theory are equal to the Bianchi identity and
the equation of motion of the T-dual theory \cite{Duff,GR,DS1,ALLP}.
The consistency conditions of the relations (\ref{eq:tdual})
\begin{equation}\label{eq:tdual1}
\partial_+  [{\cal{H}}_{MN}  \partial_- Z^N] + \partial_-  [{\cal{H}}_{MN}  \partial_+ Z^N ] \cong 0  \, ,
\end{equation}
in components take a form
\begin{eqnarray}\label{eq:em}
  \partial_+  \partial_- x^\mu \cong 0  \,  , \qquad   \partial_+   \partial_- y_\nu \cong 0 \,  .
\end{eqnarray}
They are the equations of motion for both initial and T-dual theories.

The expression (\ref{eq:tdual1}) originated from conservation of the topological currents
$i^{\alpha M}= \varepsilon^{\alpha \beta} \partial_\beta Z^M$. It is often called Bianchi identity.
In this sense T-duality  in the double space  unites equations of motion and Bianchi identities in a single relation (\ref{eq:tdual1}) as is shown in \cite{Duff}.

We can write the action
\begin{equation}\label{eq:act}
S = \frac{\kappa}{4} \int d^2 \xi \,\,  \partial_+ Z^M  {\cal{H}}_{MN} \partial_- Z^N   \,  ,
\end{equation}
which variation produce the eq.(\ref{eq:tdual1}).

\cleq

\section{T-duality as coordinates permutations in double space}

Let us mark the T-dualization along some direction $x^{\mu_1}$ by $T^{\mu_1}$, and its inverse along corresponding direction $y_{\mu_1}$ by $T_{\mu_1}$.
Up to now we collected the results from T-dualizations along all directions $x^\mu\,  (\mu=0,1,\cdots,D-1)\,$, $T^{full}=T^{0}\circ T^{1}\circ\cdots\circ T^{D-1}$
and from its inverse along all directions $y_\mu\,$ $T_{full}=T_{0}\circ T_{1}\circ\cdots\circ T_{D-1}$.
So, the relation (\ref{eq:tdual}) in fact contains T-dualizations along all directions $x^\mu$ and $y_\mu \,$
${\cal T}=T^{full} \circ T_{full}$.

In this section we will show that relation (\ref{eq:tdual}) contains information about any individual T-dualizations along some direction $x^{\mu_1}$
and corresponding one $y_{\mu_1}$ for fixed $\mu_1$ (${\cal T}^{\mu_1} = T^{\mu_1} \circ T_{\mu_1}$).
Applying the same procedure to the arbitrary subset of directions we will be able to
obtain all possible T-dualizations. It means that we are able to connect any two backgrounds in the chain (\ref{eq:chain2})
and treat all theories connected by T-dualities in a unified manner.

Let us split  coordinate index $\mu$ into $a$ and $i$  ( $a=0,\cdots,d-1$,  $i=d,\cdots,D-1$),
and perform T-dualization along direction $x^a$ and $y_a$
\begin{equation}
{\cal T}^{a}=T^a \circ T_a  \,  ,  \quad    T^a \equiv T^{0}\circ T^{1}\circ\cdots\circ T^{d-1} \,  ,  \quad
T_a \equiv T_{0}\circ T_{1}\circ\cdots\circ T_{d-1} \,  .
\end{equation}
We will show that such T-dualization  can be obtained just by exchanging places of coordinates
$x^a$ and $y_a$. Note that the  double space contains coordinates of two spaces which are totally dual relative to one another.
In the beginning these two theories are the initial one $S(x^\mu)$ and its T-dual along all coordinates $S(y_\mu)$. Arbitrary T-dualization
in the double space along $d$ coordinate with index $a, \,\, {\cal T}^{a}$,  transforms at the same time $S(x^\mu)$ to $S[y_a, x^i]$ and
$S(y_\mu)$ to $S[x^a, y_i]$. The obtained theories are also totally T-dual relative to one another.


\subsection{The coordinates permutations in double space}

Permutation of the initial coordinates $x^a$ with its T-dual $y_a$ we can realize by multiplying double space coordinate (\ref{eq:escoor}), now written as
\begin{equation}\label{eq:escoorai}
Z^M=\left (
\begin{array}{c}
 x^a  \\
 x^i  \\
y_a  \\
y_i
\end{array}\right )\, ,
\end{equation}
by the constant symmetric matrix  $({\cal T}^{a})^T = {\cal T}^{a}$
\begin{equation}\label{eq:taua}
{\cal T}^{a} {}^M {}_N = \left (
\begin{array}{cc}
1-P_a  & P_a   \\
P_a  &  1- P_a
\end{array}\right ) =  \left (
\begin{array}{cccc}
0  & 0  & 1_a & 0  \\
0  & 1_i  & 0 & 0  \\
1_a & 0  & 0 & 0  \\
0  & 0  & 0 & 1_i
\end{array}\right )\,  .
\end{equation}
Here $P_a$ is $D \times D$ projector with $d$ units on the main diagonal
\begin{equation}\label{eq:pata}
P_a = \left (
\begin{array}{cc}
1_a  &  0  \\
 0 &  0
\end{array}\right )\, ,
\end{equation}
where $1_a$ and $1_i$ are $d$ and $D-d$ dimensional identity matrices. In Ref.\cite{GPR} this transformation is called factorized duality.

Note also that
\begin{equation}\label{eq:tprop}
({\cal T}^{a}  {\cal T}^{a})^M {}_N  = \delta^M {}_N  \, , \qquad \quad   (\Omega {\cal T}^{a} \Omega)^M {}_N  = ({\cal T}^{a})^M {}_N   \, , \qquad \quad   {\cal T}^{a} \Omega  {\cal T}^{a}= \Omega \, .
\end{equation}
The last relation means that ${\cal T}^{a} \in SO(D,D)$. More precisely, we will see that ${\cal T}^{a}$  is in fact element of permutation group,
which is a subgroup of $SO(D,D)$.

We will require that the dual extended space coordinate,
\begin{equation}\label{eq:escoorai}
Z_a^M = {\cal T}^{a} {}^M {}_N Z^N =
\left (
\begin{array}{c}
 y_a  \\
 x^i  \\
x^a   \\
y_i
\end{array}\right )\, ,
\end{equation}
satisfy the same form of  the T-duality transformations (\ref{eq:tdual}) as the initial one
\begin{equation}\label{eq:tdualdf}
\partial_{\pm} Z_a^M \cong \pm \, \Omega^{MN} {}_a {\cal{H}}_{NK} \,\partial_{\pm}Z_a^K \, .
\end{equation}
Consequently, with the help of second equation (\ref{eq:tprop}) we find the dual generalized metric
\begin{equation}\label{eq:dualgm}
{}_a {\cal{H}} =  {\cal T}^{a}  {\cal{H}} {\cal T}^{a} \, ,
\end{equation}
or explicitly
\begin{equation}\label{eq:dualh0}
{}_a {\cal{H}}_{MN} = \left (
\begin{array}{cccc}
(G^{-1})^{ab}       & 2 (G^{-1}b)^a{}_j  &   2 (G^{-1}b)^a{}_b   & (G^{-1})^{aj}  \\
-2 (bG^{-1})_i{}^b  & g_{ij}             &  g_{ib}               &  -2 (bG^{-1})_i{}^j  \\
-2 (bG^{-1})_a{}^b  & g_{aj}             &   g_{ab}              & -2 (bG^{-1})_a{}^j  \\
(G^{-1})^{ib}       & 2 (G^{-1}b)^i{}_j  & 2 (G^{-1}b)^i{}_b     & (G^{-1})^{ij}
\end{array}\right )\, .
\end{equation}


\subsection{Explicit form of T-duality transformations}

Rewriting eq. (\ref{eq:tdualdf}) in components we get
\begin{eqnarray}\label{eq:comp}
\pm \partial_\pm y_a &\cong&  -2 (bG^{-1})_a{}^b \partial_\pm y_b + g_{aj} \partial_\pm  x^j + g_{ab} \partial_\pm  x^b   -2 (bG^{-1})_a{}^j \partial_\pm y_j \nonumber \\
\pm \partial_\pm x^i &\cong&  (G^{-1})^{ib} \partial_\pm y_b + 2 (G^{-1}b)^i{}_j \partial_\pm  x^j + 2 (G^{-1}b)^i{}_b  \partial_\pm  x^b  +(G^{-1})^{ij} \partial_\pm y_j \nonumber \\
\pm \partial_\pm x^a &\cong & (G^{-1})^{ab} \partial_\pm y_b + 2 (G^{-1}b)^a{}_j \partial_\pm  x^j + 2 (G^{-1}b)^a{}_b  \partial_\pm  x^b  +(G^{-1})^{aj} \partial_\pm y_j \nonumber \\
\pm \partial_\pm y_i &\cong&  -2 (bG^{-1})_i{}^b \partial_\pm y_b + g_{ij} \partial_\pm  x^j + g_{ib} \partial_\pm  x^b   -2 (bG^{-1})_i{}^j \partial_\pm y_j \, .
\end{eqnarray}
Eliminating $y_i$ from the second and third equations we find
\begin{equation}\label{eq:pard0}
\Pi_{\mp ab} \partial_\pm x^b + \Pi_{\mp ai} \partial_\pm x^i + \frac{1}{2} \partial_\pm y_a \cong 0   \, .
\end{equation}
Multiplication with  $2\kappa {\hat \theta}^{ab}_{ \pm}$, which according to (\ref{eq:inv}) is  the inverse of $\Pi_{ \mp ab}$, gives
\begin{equation}\label{eq:pard0s}
\partial_\pm x^a \cong -2\kappa {\hat \theta}^{ab}_{\pm} \Pi_{\mp bi} \partial_\pm x^i - \kappa {\hat \theta}^{ab}_{ \pm} \partial_\pm y_b \, .
\end{equation}

Similarly, eliminating $y_a$ from the second and third equations we get
\begin{equation}\label{eq:pard1}
\Pi_{\mp ij} \partial_\pm x^j + \Pi_{\mp ia} \partial_\pm x^a + \frac{1}{2} \partial_\pm y_i \cong 0   \, ,
\end{equation}
which after multiplication with  $2\kappa {\hat \theta}^{ij}_{ \pm}$, the inverse of $\Pi_{ \mp ij}$,  produces
\begin{equation}\label{eq:pard1s}
\partial_\pm x^i \cong -2\kappa {\hat \theta}^{ij}_{\pm} \Pi_{\mp ja} \partial_\pm x^a - \kappa {\hat \theta}^{ij}_{ \pm} \partial_\pm y_j \, .
\end{equation}

The equation (\ref{eq:pard0s}) is the T-duality transformations for $x^a$ (eq. (44) of ref. \cite{DNS2}) and (\ref{eq:pard1s}) is its analogue
for $x^i$.


\subsection{T-dual background fields}

Requiring that the dual generalized metric (\ref{eq:dualh0}) has the form (\ref{eq:gm}) but with T-dual background fields,
(denoted by lower index $a$ on the left of background fields)
\begin{equation}\label{eq:h0d}
{}_a {\cal{H}}_{MN} = \left (
\begin{array}{cc}
{}_a g^{\mu \nu}  &  -2  ({}_a b \,{}_aG^{-1})^\mu{}_\nu   \\
2 ({}_a G^{-1}\, {}_a b)_\mu {}^\nu  & ({}_a G^{-1})_{\mu \nu}
\end{array}\right )\, ,
\end{equation}
we can find expressions for the T-dual background fields in terms of the initial ones.

It is useful to consider the combination of the dual background fields in the form
\begin{equation}\label{eq:dppm}
{}_a \Pi_{ \pm}^{\mu \nu} \equiv ({}_a b \pm \frac{1}{2} {}_a G )^{\mu \nu}     = {}_a G^{\mu \rho}  [( {}_a G^{-1} \,{}_a b)_\rho{}^\nu \pm \frac{1}{2} \delta^\nu_\rho ] \,  \, .
\end{equation}

Comparing lower $D$ rows of  expressions (\ref{eq:dualh0}) and (\ref{eq:h0d}) we find
\begin{equation}\label{eq:bg}
( {}_aG^{-1} \, {}_a b)_\mu{}^\nu  = \left (
\begin{array}{cc}
- (bG^{-1})_a{}^b       & \frac{1}{2} g_{aj}    \\
  \frac{1}{2} (G^{-1})^{ib}    &   (G^{-1}b)^i{}_j
\end{array}\right )  \equiv \left (
\begin{array}{cc}
-{\tilde \beta}    &  \frac{1}{2}  g^T   \\
 \frac{1}{2} \gamma  &   -{\bar \beta}^T
\end{array}\right )              \,  ,
\end{equation}
and
\begin{equation}\label{eq:gmji}
({}_a G^{-1})_{\mu \nu}  = \left (
\begin{array}{cc}
 g_{ab}              & -2 (bG^{-1})_a{}^j  \\
2 (G^{-1}b)^i{}_b     & (G^{-1})^{ij}
\end{array}\right )  \equiv \left (
\begin{array}{cc}
{\tilde g}    &   -2 \beta_1 \\
-2 \beta_1^T    &   {\bar \gamma}
\end{array}\right )     \, .
\end{equation}
The notation in the second equalities which has been obtained using (\ref{eq:gnmj}), (\ref{eq:gdef}), (\ref{eq:bgnmj}) and (\ref{eq:gnmjb}) will
simplify calculations.

To obtain background field (\ref{eq:dppm}) we need the inverse of last expression.
We will use the general expression for block wise inversion matrices
\begin{equation}\label{eq:im}
\left(\begin{array}{cc}
A & B\\
C & D
\end{array}\right)^{-1}\,=
\left(
\begin{array}{cc}
(A-BD^{-1}C)^{-1}          &     -A^{-1}B(D-CA^{-1}B)^{-1}\\
-D^{-1}C(A-BD^{-1}C)^{-1}  &      (D-CA^{-1}B)^{-1}
\end{array}\right).
\nonumber\\
\end{equation}
It produces
\begin{equation}\label{eq:gmj}
({}_a G)^{\mu \nu}  = \left (
\begin{array}{cc}
 (A^{-1})^{ab}                              &   2 ({\tilde g}^{-1} \beta_1 D^{-1})^a{}_{j} \\
2 ({\bar \gamma}^{-1} \beta_1^T A^{-1})_i{}^{b}                & (D^{-1})_{ij}
\end{array}\right )\, ,
\end{equation}
where
\begin{equation}\label{eq:AD}
A_{ab}= ({\tilde g} -4  \beta_1  {\bar \gamma}^{-1} \beta_1^T)_{ab}  \, ,  \qquad \quad
D^{ij} = ({\bar \gamma} -4  \beta_1^T  {\tilde g}^{-1} \beta_1)^{ij}  \, .
\end{equation}

After some direct calculations it can be shown that
\begin{equation}\label{eq:A}
A_{ab}= ({\tilde G} -4 {\tilde b}{\tilde G}^{-1}{\tilde b})_{ab} \equiv \hat{g}_{ab}  \, ,
\end{equation}
where $\hat{g}_{ab}$ has been defined in (\ref{eq:ghat}).
Note that unlike ${\tilde g}_{ab}$, which is just $ab$ component of $g_{\mu \nu}$,
the $\hat{g}_{ab}$ has the same form as effective metric $g_{\mu \nu}$ but with all components (${\tilde G}, {\tilde b}$)
defined in $d$ dimensional subspace with indices $a,b$.

Using result (\ref{eq:A}) we can rewrite the first equation (\ref{eq:AD}) in the form
$\hat{g}_{ab}= {\tilde g}_{ab} -4 (\beta_1  {\bar \gamma}^{-1} \beta_1^T)_{ab}$.
Multiplying it on the left with $({\tilde g}^{-1})^{ab}$ and on the right with $(\hat{g}^{-1})^{ab}$ we get
\begin{equation}\label{eq:or}
({\tilde g}^{-1})^{ab} = (\hat{g}^{-1})^{ab} -4 ({\tilde g}^{-1} \beta_1  {\bar \gamma}^{-1} \beta_1^T \hat{g}^{-1})^{ab} \, .
\end{equation}
With the help of this relation we can verify that
\begin{equation}\label{eq:Dinv}
(D^{-1})_{ij} = (\bar{\gamma}^{-1} + 4 \bar{\gamma}^{-1}   \beta_1^T  \hat{g}^{-1} \beta_1  \bar{\gamma}^{-1})_{ij} \, ,
\end{equation}
is inverse of the second equation (\ref{eq:AD}).

Now, we are able to calculate background field (\ref{eq:dppm})
\begin{equation}\label{eq:pipm}
{}_a \Pi_{ \pm}^{\mu \nu}  = \left (
\begin{array}{cc}
 {\tilde g}^{-1} \beta_1 D^{-1} \gamma - A^{-1} ({\tilde \beta} \mp \frac{1}{2})         &   \frac{1}{2} A^{-1} g^T -2  {\tilde g}^{-1} \beta_1 D^{-1} ({\bar \beta}^T \mp \frac{1}{2})    \\
 \frac{1}{2}D^{-1} \gamma -2 \bar{\gamma}^{-1} \beta_1^T A^{-1} ({\tilde \beta} \mp \frac{1}{2})      &   \bar{\gamma}^{-1} \beta_1^T A^{-1}g^T - D^{-1} ({\bar \beta}^T \mp \frac{1}{2})
\end{array}\right )\, .
\end{equation}
After tedious calculations using (\ref{eq:cgama})-(\ref{eq:gnmjb}) and (\ref{eq:usr})   we can obtain
\begin{equation}\label{eq:pipmr}
{}_a \Pi_{ \pm}^{\mu \nu}  = \left (
\begin{array}{cc}
 \frac{\kappa}{2} {\hat \theta}_{ \mp}^{ab}   &  \kappa {\hat \theta}_{\mp}^{ab} \Pi_{\pm bi}  \\
-\kappa \Pi_{\pm ib} {\hat \theta}_{ \mp}^{ba}    &   \Pi_{\pm ij} -2\kappa \Pi_{\pm ia} \hat\theta^{ab}_{\mp}\Pi_{\pm bj}
\end{array}\right )\, ,
\end{equation}
where ${\hat \theta}_{ \pm}^{ab}$ has been defined in (\ref{eq:bfc}).

It still remains to check that upper $D$ rows of (\ref{eq:dualh0}) and (\ref{eq:h0d}) produce the same expressions for T-dual
background fields. The field  $({}_a b \,{}_aG^{-1})^\mu{}_\nu$ is just transpose of $({}_a G^{-1}\, {}_a b)_\mu {}^\nu$. It is useful to express
${}_a g^{\mu \nu}$ in the form
\begin{equation}
{}_a g^{\mu \nu} = ({}_a G)^{\mu \rho} [\delta_\rho^\nu -4 ({}_a G^{-1}\, {}_a b)_\rho {}^\sigma ({}_a G^{-1}\, {}_a b)_\sigma {}^\nu] \, .
\end{equation}
Then using (\ref{eq:gmj}), (\ref{eq:bg}), (\ref{eq:A}), and (\ref{eq:usr}) we can show that
\begin{equation}
{}_a g^{\mu \nu} =
\left(
\begin{array}{cc}
(G^{-1})^{ab}       & 2 (G^{-1}b)^a{}_j   \\
-2 (bG^{-1})_i{}^b  & g_{ij}
\end{array}\right )\, ,
\end{equation}
which is in agreement with (\ref{eq:dualh0}).

Consequently, we obtained the T-dual background fields in the flat background after dualization along  directions $x^a, \, (a=0,1, \cdots , d-1)$
\begin{eqnarray}\label{eq:tdualf}
& {}_a \Pi_{ \pm}^{ab} =   \frac{\kappa}{2} {\hat \theta}_{ \mp}^{ab}  \, ,  \qquad \qquad        & {}_a \Pi_{ \pm}^a{}_i =   \kappa {\hat \theta}_{ \mp}^{ab} \Pi_{ \pm bi}   \, ,  \nonumber \\
& {}_a  \Pi_{ \pm i}{}^a =  -\kappa \Pi_{ \pm ib} {\hat \theta}_{ \mp}^{ba}  \, ,       & {}_a \Pi_{ \pm ij} = \Pi_{ \pm ij} -2\kappa \Pi_{ \pm ia} \hat\theta^{ab}_{ \mp}\Pi_{ \pm bj} \,  .
\end{eqnarray}
The symmetric and antisymmetric parts of these expressions produce T-dual metric and T-dual Kalb-Ramond field.
This is in complete agreement with the Refs.\cite{DNS2,SN}. The similar way to perform T-duality in the flat space-time for $D=3$ has been described in App. B
of ref. \cite{ALLP}.

This proves that exchange the places of some coordinates $x^a$ with its T-dual $y_a$ in the flat double space represents T-dualities along
these coordinates.

 In Sec. 4.1. of Ref.\cite{GPR}  the  Buscher's T-dualities has been derived in  eq.(4.9) in the case when there is only one isometry direction. For such a case it was concluded that
 "the dual background is related to the original one by the action of factorized duality".
There is essential difference  between such  eq.(4.9) and relation (\ref{eq:tdualf}) of the present article, where the general case of T-dualityes along arbitrary sets of coordinates has been derived and proof its equivalence with the  action of factorized duality.

For proof of  expression (\ref{eq:tdualf}) with mathematical induction eq.(4.9) is just first step for $n=1$. The next step from $n$ to $n+1$ is nontrivial because then we have three kind of variables (beside isometry one $\theta$ there are a set of original variables and a set of variables along which we already performed duality transformations). This leads to the formulae different from eq.(4.9). For example,
when we performed T-dualization along more then one coordinate (lets say along $x^a, a=1,2$) in expression for T-dual background fields it is not carried out division  with $G_{a a}$ as in eq.(4.9) but with $G_{a b}+2 B_{a b}$  which was recorded in expression $\hat \theta_-^{a b}$ of (\ref{eq:tdualf}).


\subsection{T-duality group}

Successively T-dualization along disjunct sets of directions ${\cal T}^{a_1}$ and ${\cal T}^{a_2}$ will produce
T-dualization along all directions  $a=a_1 \cup \, a_2$
\begin{equation}
{\cal T}^{a_1} \circ {\cal T}^{a_2} = {\cal T}^{a} \,   .
\end{equation}
This can be represent by matrix multiplications $({\cal T}^{a_1} {\cal T}^{a_2})^M{}_N = ({\cal T}^{a})^M{}_N$, which is easy to
check because the projectors satisfy the relations $P_{a_1}^2=P_{a_1} $, $P_{a_2}^2=P_{a_2} $,  $P_{a_1} P_{a_2} =0$ and
$P_{a_1}+P_{a_2} =P_{a} $.

The set of matrices ${\cal T}^{a}$, where the index $a$ take the values in any of the subset of index $\mu$ form a commutative group
with respect to matrix multiplication. The neutral element corresponds to the case when we do not perform T-duality, with $P_a =0$ and
${\cal T}^{a}=1$. Consequently, the set of all T-duality transformations form a commutative group with respect to the operation $\circ $.

This is subgroup of the $2D$ permutational group because it acts as a replacement of some coordinates. In two-line notation, the T-duality ${\cal T}^a $,
along $2d$ coordinates $x^a \,$ and  $y_a \,$  can be written as
\begin{equation}\label{eq:pg}
 \left (
\begin{array}{ccccccccccccc}
1      & 2   & \cdots    &   d  & d+1 & \cdots & D & D+1 & \cdots &   D+d  & D+d+1 & \cdots & 2D \\
D+1   &  D+2  & \cdots  &  D+d & d+1  &\cdots & D & 1      & \cdots & d  & D+d+1 & \cdots & 2D
\end{array}\right )\, .
\end{equation}
It looks simpler in the cyclic notation
\begin{equation}\label{eq:cpg}
(1,D+1) (2,D+2) \cdots (d, D+d)   \, .
\end{equation}
We will call this group T-duality group. It is a global symmetry group of equations of motion (\ref{eq:tdual1}).


\cleq

\section{Inclusion of Dilaton field}

As usual,  in the standard formulation  one should  add Fradkin-Tseytlin term
\begin{equation}\label{eq:FTa}
S_\phi = \int d^2 \xi \sqrt{-g} R^{(2)} \phi  \, ,
\end{equation}
to the action (\ref{eq:action0}) in order to describe  dilaton field $\phi$.
Here $R^{(2)}$ is scalar curvature of the world sheet and the term $S_\phi$  is one order higher in $\alpha^\prime$ then terms in (\ref{eq:action0}).



\subsection{Path integral measure}

It is well known that dilaton transformation has quantum origin. For a constant background the Gaussian path integral produces
the expression $(\det \Pi_{ + \mu \nu})^{-1}$. We will show that this is just what we need
in order that the change of space-time measure in the path integral  is correct.

Let us start with the relations
\begin{equation}\label{eq:drg}
\det G_{\mu \nu}=  \frac{\det G_{a b}}{\det {\bar \gamma}^{i j}} \, , \qquad  \det {}_a G_{\mu \nu}=  \frac{\det {}_a G^{a b}}{\det {\bar \gamma}^{i j}}  \, ,
\end{equation}
which follow from  (\ref{eq:G}) and (\ref{eq:gmji}).
Using the expressions for T-dual fields  (\ref{eq:tdualf})   we can find the relations between the determinants
\begin{equation}\label{eq:dr}
\det (2 \Pi_{\pm ab})=  \frac{1}{\det ( 2 \, {}_a \Pi_{\pm}^{ ab})} = \sqrt{\frac{\det G_{\mu \nu}}{\det {}_a G_{\mu \nu}}}  = \sqrt{\frac{\det G_{a b}}{\det {}_a G^{a b}}}   \, ,
\end{equation}
where the  factor $2$ introduced for the convenience, because $\Pi_{\pm ab} = B_{a b}\pm \frac{1}{2}G_{a b}$.
So, we have
\begin{equation}\label{eq:ccm}
\sqrt{\det G_{\mu \nu}} \, d x^i d x^a  \to     \sqrt{\det G_{\mu \nu}}\,  d x^i    \frac{1}{\det ( 2 \,  \Pi_{+ ab})}\,  d y_a =
\sqrt{\det {}_a G_{\mu \nu}} \, d x^i d y_a \, ,
\end{equation}
which means that T-dualization $T^a$ along  $x^a$ directions produces correct change of space-time measure in the path integral of the
standard approach.


\subsection{Dilaton in the double space}

In the double space T-dualization ${\cal T}^a$ along both $x^a$ and $y_a$ produces
\begin{eqnarray}\label{eq:dim}
& \sqrt{\det G_{\mu \nu}} \,    \sqrt{\det {}^\star G^{\mu \nu}} \,d x^i    d y_i d x^a d y_a   \to   \\ \nonumber
 \\ \nonumber
& \sqrt{\det G_{\mu \nu}} \,   \sqrt{\det {}^\star G^{\mu \nu}} \,d x^i    d y_i  d y_a  d x^a
\frac{1}{\det ( 2 \,  \Pi_{+ ab}) \det ( 2 \, {}_a \Pi_{+}^{ ab})}\,    \,  .
\end{eqnarray}
According to (\ref{eq:dr}) the last term is equal to $1$ and  the  path integral measure is invariant under T-dual transformation.
Consequently, in double space we need the new dilaton  invariant under T-duality transformations.

The usual approach in the literature is to introduce "doubled dilaton" $\Phi^{(a)}$ in term of the standard dilaton $\phi$, with requirement that $\Phi^{(a)}$ is
invariant under T-dualization ${\cal T}^a$.
From the transformation of standard dilaton $\phi$ (see Refs.\cite{B,GR})
\begin{equation}\label{eq:dsh}
 {}_a \phi  = \phi - \ln \det ( 2 \Pi_{+ ab}) = \phi -  \ln \sqrt{\frac{\det G_{a b}}{\det {}_a G^{a b}}} \, ,
\end{equation}
with the help of (\ref{eq:dr}) we have $ {}_a ({}_a \phi)  = \phi$, which means that
\begin{equation}\label{eq:did}
\Phi^{(a)} = \frac{1}{2}  ({}_a \phi  + \phi) = \phi - \frac{1}{2} \ln \sqrt{\frac{\det G_{a b}}{\det {}_a G^{a b}}} \, ,
\end{equation}
is  invariant under duality transformation along $x^a$ directions.
If we chose the other set of coordinates $x^b \, (b \neq a)$,  along which we perform T-duality, then we wil have different "doubled dilaton" $\Phi^{(b)}$. We want to have one doubled dilaton invariant under all T-duality transformations.

Up to now, we described all  T-dual transformations with one action (\ref{eq:act}).
Using (\ref{eq:gm}) and (\ref{eq:cdi2}) the corresponding generalized metric
can be expressed  symmetrically in term of initial metric and Kalb-Ramond  fields and their totally T-dual background fields (marked with star)
\begin{equation}\label{eq:gmid}
{\cal{H}}_{MN}  = \left (
\begin{array}{cc}
({}^{\star} G^{-1})_{\mu \nu}    &  2 ({}^{\star} G^{-1})^{\mu \rho}  \,  {}^{\star} B_{\rho \nu}     \\
2 (G^{-1})^{\mu \rho} \,  B_{\rho \nu}  & (G^{-1})^{\mu \nu}
\end{array}\right )\, .
\end{equation}
We can do a similar thing with dilaton field. As was shown in Ref.\cite{SB}  the expression
\begin{equation}\label{eq:nd}
\Phi =  \phi - \ln \sqrt{\det G_{\mu \nu}}     \, ,
\end{equation}
is duality invariant under all possible T-dualizations.
So, the double space action (\ref{eq:act}) can be extended with the expression similar to (\ref{eq:FTa}), but with doubled dilaton $\Phi$ instead of the standard one $\phi$.

Using the fact that $\Phi = {}^\star \Phi = {}^\star \phi - \ln \sqrt{\det {}^\star G^{\mu \nu}}$,  we can express double dilaton $\Phi$
in term of dilaton from the initial theory $\phi$ and dilaton from its totally T-dual theory ${}^\star \phi$ as
\begin{equation}\label{eq:ndsm}
e^{ -2 \Phi} =   e^{- (\phi + {}^\star  \phi) } \sqrt{\det G_{\mu \nu}  \det {}^\star G^{\mu \nu}}     \, .
\end{equation}
Therefore, we can take   $e^{ -2 \Phi} d x^\mu d y_\mu$ as double space integration  measure, as well as in the Double field theory.


\section{Relation with the Hull's formulation}

In this section we are going to derive the action of Ref.\cite{Hull} and compare its consequences with our results.
Note that the background fields of Ref.\cite{Hull} depend only on the coordinates $Y^m$  ($x^i$  in our notation) along which the T-duality has not been executed. In our approach all variables are doubled $x^\mu \to y_\mu$, while in Ref.\cite{Hull} only variables  along which the T-duality is performed are doubled $x^a \to y_a$. So, in our approach there are $2D$ variables $x^a, x^i,y_a, y_i$ while in Ref.\cite{Hull} there are $D+d$ variables $x^a, y_a, x^i$. It suggest that formulation of Ref.\cite{Hull} can be obtained from our one after elimination of $y_i$ variable. We already did it in Subsec. 3.2. and obtain T-duality relations (\ref{eq:pard0}) and  (\ref{eq:pard0s}), which are inverse to each other. In analogy with (\ref{eq:tdualc}) we can rewrite them in a useful form,   where on the left hand side we put the terms with world-sheet antisymmetric tensor $\varepsilon_\alpha{}^\beta$ ($\varepsilon_\pm{}^\pm = \pm 1$) and obtain
\begin{eqnarray}\label{eq:ptdual}
\partial_\pm  X^A = \pm \, \Omega^{A B}  ({\hat {\cal H}}_{B C} \partial_\pm  X^C + J_{\pm B} )   \,  .
\end{eqnarray}
Here
\begin{equation}\label{eq:xescoor}
X^A=\left (
\begin{array}{c}
 x^a  \\
y_a
\end{array}\right )\, ,
\end{equation}
is  $2d$ dimensions double space coordinate
\begin{equation}
\Omega^{AB}= \left (
\begin{array}{cc}
0 &  1_a \\
1_a  & 0
\end{array}\right )\, ,
\end{equation}
and
\begin{equation}\label{eq:dh0}
{\hat {\cal{H}}}_{AB} = \left (
\begin{array}{cc}
{\hat g}_{ab}  &  -2  b_{ac} ({\tilde G}^{-1})^{cb}   \\
2 ({\tilde G}^{-1})^{ac}b_{cb }   & ({\tilde G}^{-1})^{ab}
\end{array}\right )\, ,
\end{equation}
is  $2d \times 2d$ generalized metric. All components of ${\hat {\cal{H}}}_{AB}$ are from $ab$ subspace, like ${\hat g}_{ab}$ and ${\hat \theta}^{ab}$ in (\ref{eq:ghat}). So,  it   satisfies ${\hat {\cal{H}}}^T \Omega {\hat {\cal{H}}} = \Omega $ and it is element of $O(d,d)$ group.

We also obtained the explicit expressions for the currents  in terms of undualized coordinates $x^i$
\begin{equation}\label{eq:curr}
J_{\pm A}=\left (
\begin{array}{c}
J_{1 \pm a}   \\
J_{2 \pm}^{a}
\end{array}\right )\, ,
\end{equation}
where
\begin{equation}\label{eq:curr12}
J_{1 \pm a}  = -2 \Pi_{\pm ab} J_{2 \pm}^{b} \,  , \qquad  J_{2 \pm}^{a}= 2 ({\tilde G}^{-1})^{ab} \Pi_{\mp bi} \partial_\pm x^i \,  .
\end{equation}
The first relation in the last expression is solution $(2.44)$ of Ref.\cite{Hull}.

Therefore, instead of $2D$ component T-duality transformations (\ref{eq:tdualdf}) with $2D$ dimensional vector $Z^M$  we have $2d$ component relation (\ref{eq:ptdual})
with $2d$ dimensional vectors $X^A$ and  $J_{\pm A}$.
The relation (\ref{eq:ptdual}) is self-duality constraints (eq.(2.5) of Ref.\cite{Hull}) imposed that halves the degrees of freedom.

As well as in Subsec. 2.3 consistency condition of (\ref{eq:ptdual}) produces
\begin{equation}\label{eq:dem}
\partial_+ ({\hat {\cal{H}}} \partial_- X + J_-) + \partial_- ({\hat {\cal{H}}} \partial_+ X + J_+) = 0   \,  ,
\end{equation}
which is equation of motion $(2.4)$ of Ref.\cite{Hull}. So, we can write the action
\begin{equation}\label{eq:dact}
S_d = \frac{\kappa}{4} \int d^2 \xi \left[\partial_+ X^A {\hat {\cal{H}}}_{AB} \partial_- X^B + \partial_+ X^A J_{-A} + J_{+A} \partial_- X^A + {\cal L}(x^i) \right] \,  ,
\end{equation}
which variation produce the eq.(\ref{eq:dem}). This action is in complete agreement with  Ref.\cite{Hull}, but with already  constrained elements of ${\hat {\cal{H}}}_{AB}$
and explicit expression for currents $J_{\pm A}$ in terms of undualized coordinates $x^i$.
Because Ref.\cite{Hull} starts with arbitrary ${\hat {\cal{H}}}_{AB}$ it is restricted to be coset metric $O(d,d)/O(d)\times O(d)$ that the T-duality would be
equivalent to the standard one.

Note that the whole procedure of the Ref.\cite{Hull} should be performed for each node of the chain (\ref{eq:chain2}), which means for each values of $d$. In our approach only the
case $d=D$ appears. Then the currents $J_{\pm A}$ disappear, $X^A \to Z^M, \,\, {\hat {\cal{H}}}_{AB} \to {\cal{H}}_{MN},\,\, \Omega^{AB} \to \Omega^{MN}$ and T-duality transformations
(\ref{eq:ptdual}) turns to (\ref{eq:tdual}).
However, the generalized metric ${\cal{H}}_{MN}$ together with basic relation (\ref{eq:tdual}) are sufficient to
describe all T-dualities for each $d$.


\section{Conclusion}

Introducing the $2D$ dimensional  space, which beside initial  $D$ dimensional space-time coordinates $x^\mu$ contains
the corresponding T-dual coordinates $y_\mu$, we offered simple formulation for T-duality transformations.
The extended space with the coordinates $Z^M= (x^\mu, y_\mu)$ we call double space.

It is easy to see that after the exchanges of all initial coordinates $x^\mu$ with all T-dual coordinates $y_\mu$ we obtain the T-dual background fields
of Sec.2. This result is formulated in the double space in Ref.\cite{Duff} in order to make global $SO(D,D)$ symmetry manifest.
In the present article we show that the double space contains enough information to explain T-dualization along arbitrary subset of
coordinates $x^a$ and corresponding T-dual $y_a$  $(a=0,1, \cdots ,d-1)$. For this purpose we rewrite
T-duality transformations for all the coordinates and its inverse in the double space. We obtain the basic relation (\ref{eq:tdual})
with the generalized metric  (\ref{eq:gm}) which help us to find all T-dual background fields for each node of the chain (\ref{eq:chain2}) and T-duality transformations between all the nodes.

We define particular permutation of the coordinates realized by matrix  ${\cal T}^a$, known in literature as factorized duality (see for example \cite{GPR}). It exchanges the places of some subset of the coordinates $x^a$ and
the corresponding dual coordinates $y_a$ along which we perform T-dualization.
We require that the obtained double space coordinates satisfy the same form of T-duality transformations as the initial one,
or in other words that such permutation is a global symmetry of the T-dual transformation.
We show that this permutation produce exactly the same T-dual background fields and T-duality transformations as
in the standard approach of Ref.\cite{DNS2}.
So, double space approach clearly explains that T-duality is nonphysical, because it is equivalent to the permutation
of some coordinates in the double space.

In the standard formulation T-duality transforms the initial theory to the equivalent one, T-dual theory. The double space formulation contains
both initial and T-dual theories and T-duality becomes the global symmetry transformation. With the help of (\ref{eq:dualgm})
it is,  easy to see that equations of motion (\ref{eq:tdual1}) are invariant under transformation
$Z^M \to Z_a^M = ({\cal T}^a )^M{}_N Z^N$.

The square of all matrix ${\cal T}^a$ are equal to one and therefore they are inverse themselves. The set of all ${\cal T}^a$ matrices form an Abelian group with respect to the matrix
multiplication. Consequently, the set  of all T-dualizations with respect to the successive T-dualizations also form an Abelian group.
It is a subgroup of the $2D$ permutation group, which permute some of the first $D$ coordinates with corresponding last $D$ coordinates. In the cyclic form
it can be written as
\begin{equation}\label{eq:cpg1}
(1,D+1) (2,D+2) \cdots (d, D+d)   \, ,   \qquad  d =  0,1,2, \cdots ,D \,  ,
\end{equation}
where  $d=0$ formally corresponds to the neutral element (no permutations of coordinates and so no T-duality transformations) and $d=D$ corresponds to the case when T-dualization is performed along all coordinates.

The relation between our approach and the well known one of Ref.\cite{Hull} has been presented in Sec.4.  In approach of Ref.\cite{Hull} to each node of the chain (\ref{eq:chain2}), lying $d$ steps from the begining, it corresponds the action $S_d$ (\ref{eq:dact}) and self-duality constraints (\ref{eq:ptdual}) with $2d$ dimensional variables $X^A$. Our approach unify all nodes of the chain (\ref{eq:chain2}). The T-duality transformations (\ref{eq:tdual}), with $2D$ dimensional variable $Z^M$, allows us to obtain all background fields and T-duality transformations of the chain (\ref{eq:chain2}).

Let us briefly describe the significance of the obtained results.
It is well known that there are five consistent  superstring theories. In order to have the unique theory, the so-called M-theory, we should connect these five  theories by web of T and S dualities.
If we start with arbitrary of these  five consistent  theories  and find all corresponding T-dual and S-dual theories we can achieve  any of other four consistent  superstring theories.
But it is not enough for formulation  of M-theory. To realize  this we should  construct one theory which contain the initial theory and all corresponding dual ones.

The present article is realization of such program for T-duality  in the bosonic case for flat background, which is substantially simpler that supersymmetric  one.
In fact, the theory with all doubled  coordinates contained the initial and all corresponding T-dual theories.
We  hoping that   S-duality, which can be understood as transformation of dilaton background field, can be  successfully  incorporated  into our procedure.
The same program for bosonic string but in the weakly curved background, with  linear  dependence   on coordinates, will be investigated in Ref.\cite{SB}.

Unfortunately, the solution for bosonic case  is not enough for construction of M-theory, because the T-duality for superstrings is non-trivial extension of the bosonic case.
In Ref.\cite{bosdouble} we have tried to extend such approach to the type II theories. In fact, doubling all bosonic coordinates we have unified types IIA, IIB as well as type $II^\star$ \cite{timelike}
(obtained by T-dualization along time-like direction)  theories.

We expect that, in  our approach to the formulation of M-theory   we should  include  T-dualization along fermionic variables, also.
It means that we should doubled these fermionic variables, also. A necessary step for understanding T-dualization along all fermionic coordinates in fermionic double space has been considered in Ref.\cite{NS} . We expect that final step in construction of M-theory will be unification of all theories obtained after T-dualization along all bosonic and all fermionic variables. In that case we should  double all coordinates in superspace, anticipating that  some
super permutation will connect arbitrary two of our five consistent super symmetric string theories.

\appendix
\cleq

\section{Block-wise expressions for background fields}\label{sec:dodatak}

In order to simplify notation and to write expressions without indices (as matrix multiplication) we will introduce notations for
component fields.

For the metric tensor and the Kalb-Ramond background fields we define
\begin{equation}\label{eq:G}
G_{\mu \nu} = \left (
\begin{array}{cc}
{\tilde G}_{ab}    &    G_{aj}       \\
 G_{ib}            &   {\bar G}_{ij}
\end{array}\right )   \equiv
 \left (
\begin{array}{cc}
{\tilde G}    &    G^T       \\
 G            &   {\bar G}
\end{array}\right )   \, ,
\end{equation}
and
\begin{equation}
b_{\mu \nu} = \left (
\begin{array}{cc}
{\tilde b}_{ab}    &    b_{aj}       \\
 b_{ib}            &   {\bar b}_{ij}
\end{array}\right )   \equiv
 \left (
\begin{array}{cc}
{\tilde b}    &    -b^T       \\
 b            &   {\bar b}
\end{array}\right )   \, .
\end{equation}
We also define notation for inverse of the matric
\begin{equation}\label{eq:gnmj}
(G^{-1})^{\mu \nu} = \left (
\begin{array}{cc}
{\tilde \gamma}^{ab}    &    \gamma^{aj}       \\
 \gamma^{ib}            &   {\bar \gamma}^{ij}
\end{array}\right )   \equiv
 \left (
\begin{array}{cc}
{\tilde \gamma}    &    \gamma^T       \\
 \gamma          &   {\bar \gamma}
\end{array}\right )   \, ,
\end{equation}
and for the effective metric
\begin{equation}\label{eq:gdef}
g_{\mu \nu} = G_{\mu \nu} -4 b_{\mu\rho} (G^{-1})^{\rho \sigma} b_{\sigma \nu}
= \left (
\begin{array}{cc}
{\tilde g}_{ab}    &    g_{aj}       \\
 g_{ib}            &   {\bar g}_{ij}
\end{array}\right )   \equiv
 \left (
\begin{array}{cc}
{\tilde g}    &    g^T       \\
 g            &   {\bar g}
\end{array}\right )   \, .
\end{equation}

Note that because $G^{\mu \nu}$ is inverse of $G_{\mu \nu}$  we have
\begin{eqnarray}\label{eq:cgama}
& \gamma = - {\bar G}^{-1} G {\tilde \gamma} = - {\bar \gamma} G {\tilde G}^{-1} \, , \qquad
& \gamma^T = - {\tilde G}^{-1} G^T {\bar \gamma} = - {\tilde \gamma} G^T {\bar G}^{-1} \, , \nonumber \\
& {\tilde \gamma} = ({\tilde G}- G^T {\bar G}^{-1} G)^{-1} \,  ,
& {\bar \gamma} = ({\bar G}- G {\tilde G}^{-1} G^T)^{-1} \,  , \nonumber \\
&{\tilde G}^{-1}= {\tilde \gamma}- \gamma^T {\bar \gamma}^{-1} \gamma \, ,
&{\bar G}^{-1}= {\bar \gamma}- \gamma {\tilde \gamma}^{-1} \gamma^T \,  .
\end{eqnarray}

It is also useful to introduce new notation for expressions
\begin{equation}\label{eq:bgnmj}
(bG^{-1})_\mu{}^\nu  = \left (
\begin{array}{cc}
{\tilde b}  {\tilde \gamma}- b^T \gamma      &    {\tilde b}\gamma^T- b^T {\bar \gamma}         \\
  b  {\tilde \gamma} + {\bar b} \gamma       &    b \gamma^T + {\bar b} {\bar \gamma}
\end{array}\right )   \equiv
 \left (
\begin{array}{cc}
{\tilde \beta}    &    \beta_1       \\
 \beta_2            &   {\bar \beta}
\end{array}\right )   \, ,
\end{equation}
and
\begin{equation}\label{eq:gnmjb}
(G^{-1}b)^\mu{}_\nu  = \left (
\begin{array}{cc}
 {\tilde \gamma} {\tilde b} + \gamma^T b      &   - {\tilde \gamma} b^T + \gamma^T {\bar b}         \\
 \gamma  {\tilde b} + {\bar \gamma} b       &    -  \gamma b^T + {\bar \gamma} {\bar b}
\end{array}\right )   \equiv
 \left (
\begin{array}{cc}
-{\tilde \beta}^T    &    -\beta_2^T       \\
 -\beta_1^T            &  - {\bar \beta}^T
\end{array}\right )   \, .
\end{equation}

We denote by $\,\,{\hat {} }\,\,$  expressions similar to  the effective metric (\ref{eq:gdef}) and non-commutativity  parameters
 but with all contributions from $ab$ subspace
\begin{equation}\label{eq:ghat}
{\hat g}_{ab} = ({\tilde G} -4 {\tilde b}  {\tilde G}^{-1} {\tilde b})_{ab}   \, , \qquad
 {\hat \theta}^{ab} = -\frac{2}{\kappa} ({\hat g}^{-1} {\tilde b} {\tilde G}^{-1})^{ab}   \, .
\end{equation}
Note that ${\hat g}_{ab} \neq {\tilde g}_{ab}$ because ${\tilde g}_{ab}$ is projection of $g_{\mu \nu}$ on subspace $ab$.
It is extremely useful to introduce background field combinations
\begin{equation}\label{eq:bfc}
\Pi_{\pm ab}= b_{ab} \pm  \frac{1}{2} G_{ab}   \,  \qquad
{\hat \theta}^{ab}_\pm =  -\frac{2}{\kappa} ({\hat g}^{-1} {\tilde \Pi}_\pm {\tilde G}^{-1})^{ab} =
{\hat \theta}^{ab} \mp \frac{1}{\kappa}  ({\hat g}^{-1})^{ab} \, ,
\end{equation}
which are inverse to each other
\begin{equation}\label{eq:inv}
{\hat \theta}^{ac}_\pm  \Pi_{\mp cb} = \frac{1}{2 \kappa} \delta^a_b   \, .
\end{equation}

With the help of  (\ref{eq:or})  one can prove the relation
\begin{eqnarray}\label{eq:usr}
({\tilde g}^{-1} \beta_1  D^{-1})^a {}_i= ({\hat g}^{-1} \beta_1 {\bar \gamma}^{-1})^a {}_i  \, ,
\end{eqnarray}
where $D^{ij}$ is defined in (\ref{eq:AD}).


\end{document}